# TEACHING CHEMISTRY IN A SOCIAL LEARNING ENVIRONMENT: FACING DRIVERS AND BARRIERS

**Cornélia Castro, António Andrade**

*School of Economics and Management, Portuguese Catholic University (PORTUGAL)*
*corneliacastro@gmail.com, aandrade@porto.ucp.pt*

## Abstract

The Portuguese Technological Plan for Education (TPE) was established to modernize schools and to consolidate the role of Information and Communication Technologies (ICT) in order to promote the academic success of students and allow schools to be transformed into technological enhanced environments through a significant learning and knowledge building in a participatory, collaborative and sharing logic.

With this work we aimed to establish dynamical interactions students-content-teacher in order to overcome a diagnosed students' lack of effort towards studying curriculum chemistry content.

Our methodology design is a non experimental and descriptive one, carried out in a secondary school during a whole school year, in order to answer the question "*How to improve the engagement of K-12 students in chemistry classes?*". Students, gathered in small groups, were asked to create digital learning resources (DLR) during classes. The teacher assumed the role of the supervisor, coach and facilitator of every task that had to be taken or chosen by the students. To enhance interaction student-student and student-teacher, a *Twitter* account and a *Ning* community were created for the class. Both supported the Social Learning Environment (SLE) that was intended to be created. The data collected – DLR created, participation of each student in the *Ning* community, teacher observations and students' opinions – led us to satisfactory results in what concerns the goals of this exploratory study.

We based our research in the following facts: i) the development of a technological basis in education is a social demand and therefore its adoption in a systemic way is no longer a matter of choice; ii) our "*digital native*" students have different learning styles not considered by traditional instructional practices; iii) SLE promotes and supports collaborative learning through the use of student-created content; iv) involving innovative pedagogical approaches opens up opportunities to engage students in creative learning and can possibly help bridge gaps between innovative and conventional pedagogical approaches and v) the use of well-structured technology, in certain contexts, may have a significant and positive effect on students' achievements.

SLE has been shaped by experience and this study is about that: web 2.0 tools, social media and student-generated content were used to answer to a combination of curriculum very specific skills and the needs and desires of the students.

The affordances and constraints of SLE as an open architecture that has potential to facilitate collaborative learning are delineated.

The methodology followed out allowed an interaction from many to many and an increased students' engagement in the proposed tasks. SLE developed ICT skills and established itself as a cognitive tool: students accessed information about learning content and turned it into knowledge.

The results seem to demonstrate that TPE, by allowing the ubiquity of computers in schools, let students to be more engaged in working and came out to be a facilitator for integrating ICT in the curriculum.

Our findings are important for the Portuguese teachers who still withstand, before these "*Net Generation*" students, to experimenting pedagogical approaches that integrate ICT into real teaching contexts which is now an imperative cause.

Future work should focus on mechanisms that allow assessment both of the methodology used and the students' generated content in order to improve students' learning in this environment.

# 1 THEORETICAL FRAMEWORK

A XXI st century decade has already end. We know that since the last century 80s, we have been living immersed in a society increasingly dominated by the so called "*new technologies*". People that were born after then, which include our students of today, grew up in a digital world very different from the one of their teacher generations. Mark Prensky [1] identified these new generations as *digital natives* [and their teachers as *digital immigrants*] and David White [2] as *residents*, taking into account the fact that students spend a long time online. Prensky [1] argues that it must then occur a radical adaptation of teaching methods to this new reality where generations with different digital expertise live and coexist. In literature, there are authors who argue, however, that the digital skills revealed by our students today are nor deep or uniform, and therefore, advocate the need to develop research in this area [3].

Aware of this reality, in this *first of all crafts* [4], we attend and participate in peer discussions, national and internationally, that the Internet and the web allow today and adopt a certain position towards these issues because, as educators, we should be reflective and not only consumers of curriculum.

We share thus the vision of Mark Prensky [5]:

> *(…) If you are an experienced teacher, you almost certainly have students filling up your classes who are, in many ways different from those in the past. You probably feel a need, or some pressure (and may have even started) to do something different for them. (…). Yet many of the teaching techniques you once used successfully do not seem to be working with today´s students. You have probably wondered about and perhaps already begun, making changes to the ways you previously taught (…).*
>
> Location 354-66, Kindle edition

The subject of chemistry, of optional nature, is part of the syllabus of the course Science and Technologies of the 12$^{th}$ grade, the last one in the secondary school in Portugal. The curriculum of the course allows "a free choice for tasks, exploring of strategies and teaching methodologies according to the students' interests and development" [6, p. 2]. It is indicated that the construction of knowledge must be framed in a broad range of skills considering how crucial it is the active involvement of students and the existence of means such as facilities, equipment, teaching resources and technical support for the development of those skills by pupils. The curriculum authors also refer that the assessment tools should take into account the evaluation of the level of performance that learning will allow to reach [6].

It seems thus clear that the chemistry teacher, *digital immigrant*, should adjust himself to the challenges posed by these "new" students, should not be a mere consumer of curriculum but a manager of it and he may, therefore use methodologies that lead the students to no longer be passive consumers of information but adopt instead an attitude of active co-construction of their knowledge. To achieve such a goal, students should mobilize (also) their digital skills – underpinned by teachers – once it will make them more engaged and committed in the classroom work and so more talented, which may be a guarantee of a better future.

Since the last decade of the twentieth century, researchers have been debating the value and effect of technology on elementary and secondary education, assuming that a proper use, in certain contexts, may have a significant and positive impact on student learning. On the other hand, several comprehensive studies have concluded that computers have a low or negative effect on student learning [7, 8]. However, a number of meta-analysis on a large scale indicated a significant improvement either in school grades or in students' attitudes towards learning and understanding when computers were integrated into the learning process. The added value observed in these studies depended, however, on several factors such as the curriculum area, the specificity of the student population, the role of the educator or the level of education [8]. The literature also highlights a number of other constraints that hinder the integration of ICT in the classroom with success: the philosophy of the school board regarding the use of technology, technological skills of teachers, fear of technological problems, lack of time, the lack of a clear knowledge of how to integrate technology into the teaching and learning process and short access to technology. All these hinderances end up explaining that even newly qualified teachers feel unprepared to use technology [8].

Despite these contradictory results in what concerns the effectiveness of technology in primary and secondary levels of schooling, educational policy experts have been making a concerted effort to increase the presence of technology in the classroom [8], of which studies of European Schoolnet European Commission [9] are an example.

The transformation of Portuguese schools and classrooms in technology enhanced environments due to the politics of implementation of the Technological Plan for Education (TPE), making computers ubiquitous in our schools, have created conditions for the reconception of the teacher's activity as a curriculum designer, now facing up the need to transform his way of teaching and of engaging students in new ways of learning.

So, school must find ways to get the maximum benefit of technology for student learning as the pedagogical arguments centred on the role of ICT in teaching and learning have become very influential as the tools become more available and useful [10].

In Science education, ICT have been established as relevant tools for these purposes, despite the views of critics above mentioned. According to Jonassen [11], students do not learn from ICT but thinking, with or without ICT. The literature also shows that teachers and their pedagogical approaches are crucial elements that can make a difference. It is therefore inevitable to consider appropriate ways of incorporating ICT and use them specifically and profitably. With this study we intended to contribute to show, not how to use ICT as a support or complement to the lessons in order to replicate the most common practices in the classroom, but rather how to use ICT in a way that allows to redesign content, objectives and the very pedagogical approach in itself [10].

In this *hiperflaction of time* we cannot continue driving into the future in a ox-cart because the wheels would fall off. Time flows faster, and the vehicle that we will have to drive must be very responsive and quick in that response because there will be many surprises and we do not know the map of what is ahead [12]. "Unlearning obsolete routines is the secret to a long life" [12, p. 2].

*Social media*, also known as *web 2.0*, relies on web-based technologies and, together with the *social tools*, facilitates the creation, sharing and interaction with information. In *web 2.0* information flows in multiple directions and in such a manner that participation becomes as important as its use [13, 14]. However, although this phenomenon is widespread by Internet users, its advancement in the educational context faces several challenges, in terms of both pedagogical and organizational aspects of the school and of technological requirements [15] and there are no ready-made solutions. This way of learning, also called *Learning 2.0* is being built on the basis of experimentation to answer to a specific combination of skills, needs and wishes of a specific group of actors [15].

## 2   METHODOLOGY

We thought of creating a Social Learning Environment (SLE) based on the assumption that it could foster the development of innovative practices in the education domain: teacher would become a coordinator, moderator and mentor – rather than just an instructor – whereas students would have to assume a pro-active role in the learning process, develop their own – individual and collective – rules and strategies for learning and jointly create content [15]. To achieve these aims we designed this descriptive – it is reported how the social learning environment was created and operationalized and non experimental study, once the phenomena observed – impressions, actions, the performance of students – such as they occur in the context of the classroom are analyzed afterwards [16].

Confident that with the available technology we would be able to develop in students skills to create and innovate, we chose to experience and to reflect on the impact on the classroom situation because, not being able to anticipate we should investigate.

We established as research question "*How to improve the engagement of K-12 students in chemistry classes*?" being our population comprised by chemistry students of the $12^{th}$ grade of a secondary school. The path tested and described below took place during a whole school year and the data were collected from a sample of sixteen students who, in previous years, had belonged to different classes with different teachers of chemistry and were now together in the same class, in this optional subject. With respect to sampling, our methodology is a convenient (or non-probabilistic) one because the choice of units was the product of peculiar circumstances [17], of the context (there was a single K-12 chemistry class in the school) and of the time in which the various elements of the sample were addressed.

The teacher that carried out this study had previously taught nine of these sixteen students during the first two levels of the upper secondary school and knew about their lack of motivation and commitment towards studying chemistry. In addition, students with the best scores in the two previous years (the other seven) and who had been taught by other teacher were diagnosed with lack of skills in laboratory work. The K-12 chemistry curriculum had to be thus administered, at the same time, to two sets of students very heterogeneous in terms of motivation for study, laboratory skills and future career ambitions.

Despite the identified drivers: i) digital skills of most of the pupils; ii) one computer in the classroom; iii) availability of broadband; iv) *e-school* government program[1], barriers were also identified: i) lack of chemical reagents and equipment for conducting laboratory classes provided in the official program of the course (the request of the missing material was carried out on time but the school principal decided not to consider it at the appropriate time); ii) toxic materials and faulty chemistry laboratory exhaustion; iii) inadequate size of the chemistry lab for the number of students; iv) heterogeneity of the class (knowledge, motivation, work, future goals) and v) problems with access to *Moodle* at school.

## 2.1 Challenges

Facing these drivers and barriers the teacher risked a paradigm shift: instead of conventional and masterful classes, with contents being taught with electronic presentations (*PowerPoint*), videos, animations and simulations followed by problem solving group activities, the following challenges were proposed to the students: i) diagnosis of their learning styles; ii) constitution of groups with two or three elements; iii) creation of a digital learning resource (DLR) of about 10-15 minutes using *web 2.0* tools in order to mobilize the students' digital skills and iv) presentation of the DLR to the class on a scheduled date.

The students answered two questionnaires on learning styles: the Index of Learning Styles Questionnaire (ILS) and the Learning Styles Questionnaire (LSQ), from a web application [18] that, in addition to the computerized automatic processing of questionnaires, allowed the teacher to prepare a report of the profile of students in the class. Based on the answers to the questionnaire, it was possible to the teacher to verify if the defined SLE was suited for the class. By consensus, the students chose their group as well as the topic to develop from among several proposed by the teacher and it was agreed that the DLR would be one among several instruments for assessing student learning. The teacher established, in each term, a dead line to finish the DLR as well as the dates for its presentation and discussion in class.

## 2.2 Operation

The students, although holders of some expertise in digital skills like messaging, youtubeing, downloading and gaming, did not know what the *web 2.0* concept meant.

Facing this scenario and with the aim to establish a social learning environment, to meet the students' learning styles, needs and wishes a class *Twitter* account – with protected tweets – and a private *Ning* account were created once we strongly believed that this open architecture would facilitate collaborative learning and would make possible the debate about DLR to present, being *Twitter* a channel for students to express their opinions.

As there was only one computer (the teacher's desk one) in the classroom, the school laptops were previously requested and when not available students would bring to class their own *e-school* program laptops. In both situations, in all laptops, Internet broadband had to be set during the course of lessons. The teacher provided the electric extension cords to connect the laptops to electricity in every single class.

The *Ning* account was created with the idea of sharing the files that were being drawn up for each DLR through an individual student's *Box.net* and with the goal to establish itself as a community of practice in which students and teachers could meet to share materials relevant to the learning of all.

---

[1] A government program aimed to provide equipment and software to – among other identified groups – students and teachers of primary and secondary education [19].

## 3 RESULTS

Despite the fact that the *Ning* account had been created with the aim cited before, it was found that students preferred to share files among each group members through *Windows Live Messenger*. The *Ning* community was then used for sharing other activities and instruments between the class and the teacher and among students, for delivery of homework, to provide support material by the teacher, for dissemination of the photographs on the activities in the classroom, to develop forums on topics initiated by the teacher and to share everything students wanted to (personal photos, news, videos or music) after teacher approval. The *Ning* community (Fig. 1) has provided therefore a contribution to the creation of the SLE and established itself as a complement to the interaction performed via *Twitter* (Fig. 2).

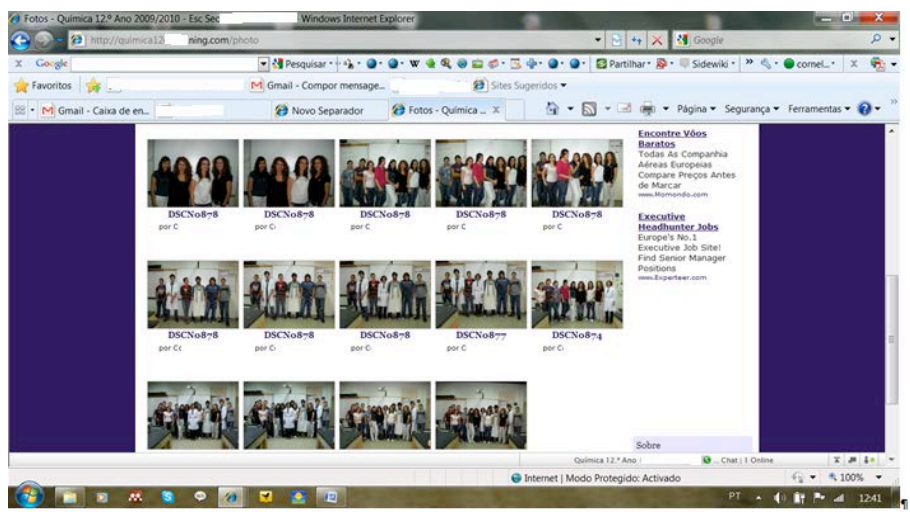

Fig. 1 – Class photos on *Ning* community (chemistry lab)

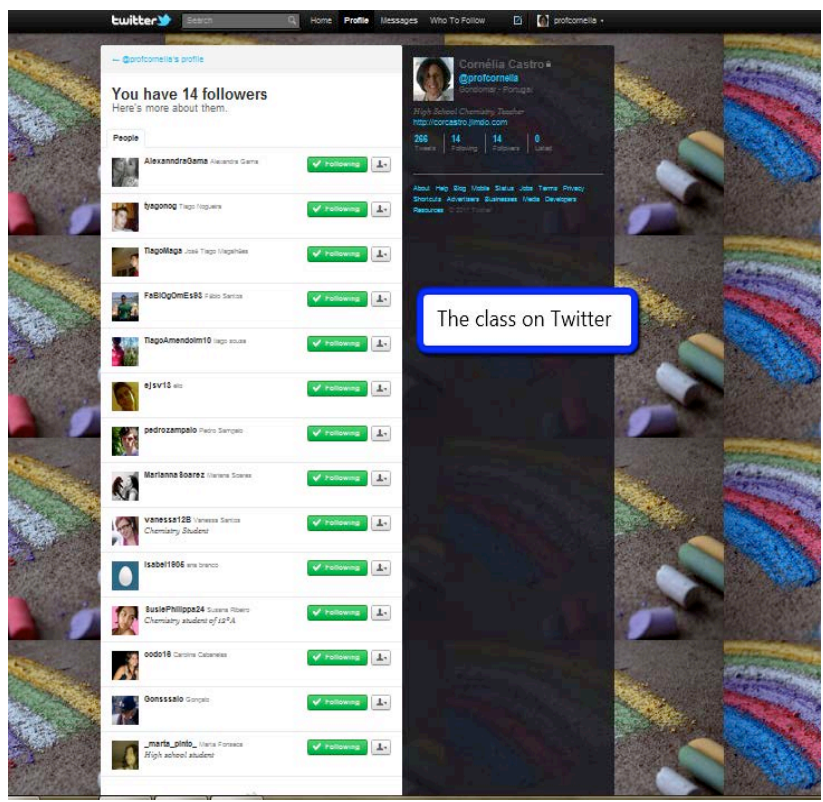

Fig. 2 – The class on *Twitter*

In all the classes where laptops were used to create a DLR, the teacher assumed the role of the supervisor, coacher and facilitator of every task that had to be taken or chosen by the students: approved the schema of the overall DLR designed by each group of students, answered all their doubts in what concerned chemistry content or which *web 2.0* tool to use, for example, but students were allowed, motivated, impelled and driven to search or "discover" other *web 2.0* tools that met or were suited to what was intended to create.

Students resorted to screen capture software, audio, *Movie Maker* and other editing tools, *Youtube*, *Vimeo*, *Flickr*, *Toonlet* (Fig. 3), *Go!Animate* (Fig. 4), *Bitstrips*, *PowerPoint*, *Creately* (Fig. 5) and *Prezi* (Fig. 6), for example.

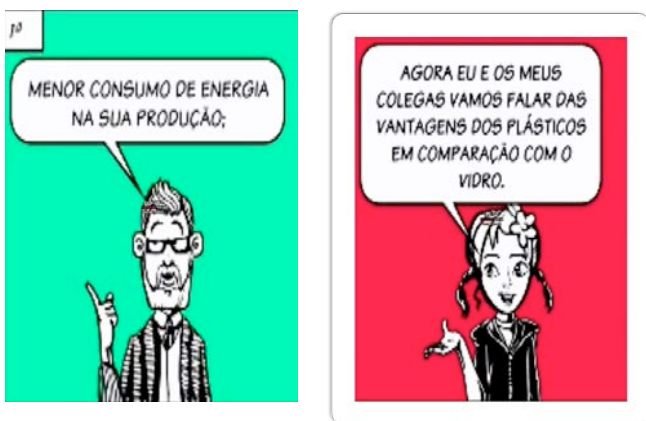
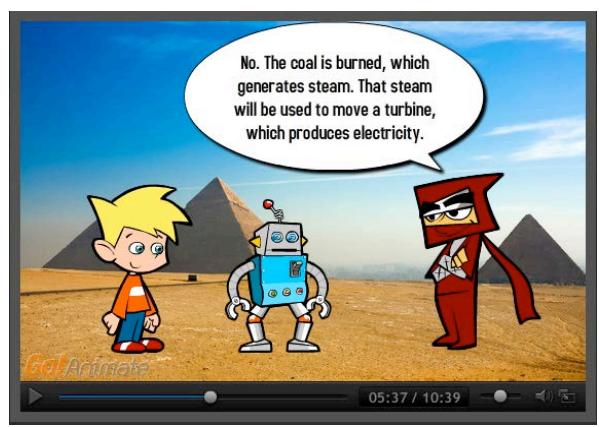

Fig. 3 – *Toonlet* strips

Fig. 4 – Chemistry content in *Go!Animate* [20]

Each of the six groups has created its DLR on the time scheduled. It was noticed that the quality of the DLR improved from the first to the last term of the school year.

The "peer-review" carried through by all the students of the class and the debates/discussions that followed the presentation of each DLR to the class may have been a contribution to this improvement in quality, in terms of depth of the content, the tools used and even the aesthetical quality of the DLR.

It is our perception that the students' commitment and motivation to study the chemistry contents proposed have improved (signs like yawns were over) as all of the students reported and remarked that the time of the classes seemed to be faster (90 minutes for the lecture classes and 135 minutes for the lab sessions). At the same time the number of doubts and questions posed by the students increased. On the other hand, *Twitter* eased communication of the shy students who used the 140 characters of this micro blogging tool to comment and assess the DLR of their colleagues.

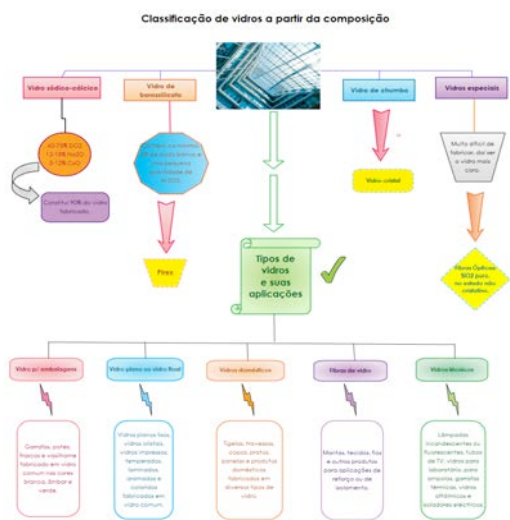
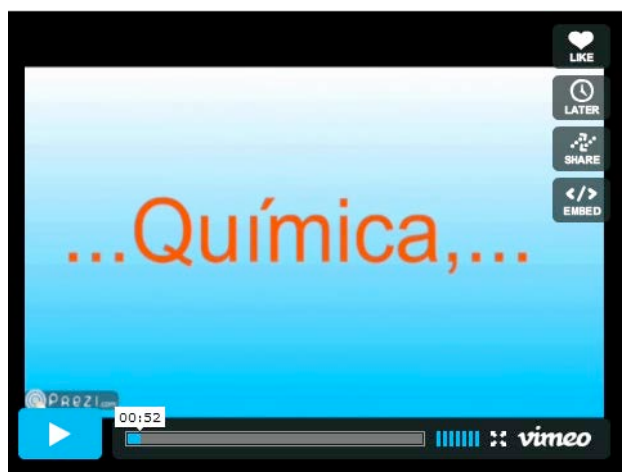

Fig. 5 – Using *Creately* to create concept diagrams

Fig. 6 – A video where *Prezi* was incorporated [21]

At the end of each term, students were asked to evaluate the procedure adopted in the classroom and some of those students' opinions were the following:

A - "This year I learned how to make better multimedia work and met some new work tools. At this point, I think I'm much more able to select the important information and neglect the irrelevant one, it´s easier to sum up the content to learn and I feel much more autonomous on studying. Working in a group taught me how to reach a unanimous decision, which is not an easy task. I learned that teachers do not necessarily have to be boring, teaching only the subjects because there are other skills that we need to develop. Someone has to teach us (…)" (female student, 17 years old).

B - "At first we were not very open to new ideas because we were not accustomed to these new methods and this has conditioned our work a bit because we thought it would be something remote and too laborious. But when we begun to embrace these new technologies, with the teacher´s encouragement, our interest increased as well as our desire to create something new and different. The teacher gave us freedom to accomplish the work in our own way and her attitude somehow made us more receptive to this new method of teaching and learning. We have tried to be creative in our work and strived utmost. (…) At this point we prefer 1000 times working with ICT as it gives us pleasure, we can achieve better results and learn the subjects with more fun" (one female student, 18 years old, two male students, 17 and 18 years old).

C - "Ouch! After all, what is it to talk about? I have not received any information on the *Ning* or *Twitter*! And this is the way how students from classes X and Y contact between them and with the teacher CC. The use of *web 2.0* tools in teaching chemistry allows us to stay abreast of what's in this virtual world and increases socialization among colleagues, students and teacher. Giving us the opportunity to work with ICT in the classroom helped and contributed to our creative development. As we were doing the work, we learned how to select important information (...) and we think that this will help us in our academic future. We started this school year making short films with *Movie Maker* and our expertise was encouraged with *Go!Animate*, for now..." (three female students, 17 years old).

The students were assessed on the DLR created, which was only one instrument among some others (already used in the two previous years) but learning and working under the SLE developed made their scores/grades become better.

## 4   CONCLUSIONS

The barriers that we faced at first i) lack of enough computers for all the students; ii) lack of broadband in some classes; iii) lack of principal support before the request of our changing to a classroom with about ten computers; iv) lack of students' knowledge towards *web 2.0* tools; v) initial students' resistance to work in a non conventional way, were overtaken by the drivers we were able to identify and develop: i) students' digital skills; ii) notorious increased students' engagement in the proposed tasks; iii) students' higher levels of responsibility in ensuring the presence of the necessary number of laptops, bringing their own as well as their *usb* internet connections.

We succeeded in shaping a Social Learning Environment based on experiencing *web 2.0* tools, *Twitter* and *Ning* accounts and student-generated content that allowed answering to a combination of curriculum very specific skills and at the same time meeting students' wishes. Their testimonies show that they enjoyed the collaborative learning and the teacher remarked that the traditional copy and paste from the Internet ended to give way to copy and… create.

The social learning environment developed ameliorated students' ICT skills and established itself as a cognitive tool once students were able, before all the information they managed to gather, to turn it into knowledge. Dynamical interactions student-content-teacher were established and the initial diagnosed students' lack of effort was overcome: students' engagement in the proposed tasks increased, term after term, they worked in a participatory, collaborative and sharing logic and achieved higher scores at the end of each term. This has been considered a very significant and positive outcome.

We consider the DLR as user(learner)–generated content, once according to OCDE [22] it is content that was made publicly available over the internet (on *Youtube*, *Vimeo* and the teacher website) and reflects a certain amount of creative effort. This kind of content was not developed or authored by experts of a domain but by learners [22] and so we thought of a way to validate the quality of the content produced. In order to achieve this, after the end of the school year, the teacher submitted one of the DLR to a Portuguese repository of Science digital learning resources. After a peer-reviewing

003383

process, the DLR was published in the repository under the metadata chemistry, K-12 and multimedia and was nominated for the annual monetary prize that the repository [23], hold by FCG – a private institution of public utility whose statutory aims are Art, Charity, Science and Education –, carries out each year [24].

We consider that support of the school principal would have been a key factor for not wasting time in preparing the classroom and computer equipment as the time spent in these tasks could have been channelled and used to perform more activities and more effective learning.

With the plethora of computers that the TPE placed in schools, it was possible to work the chemistry content so as to achieve a greater commitment of our twenty-first century students in the classroom work and the effect on their achievements in learning chemistry was considered positive and significant.

It was thus possible to answer the initial research question with the results from this study and we believe that the teaching approach implemented can possibly help bridge gaps between innovative and conventional pedagogical approaches.

Nevertheless, the TPE in itself can only be considered an important factor if the teacher, as an education professional, keeps in mind that knowledge can only be built in context and from experience and so educators need to overcome fearfulness or professional inertia, think about and mobilize other logics of action, besides the conventional ones in what concerns the teaching and learning process.

The individual role of the educator is therefore essential in order to meet the diverseness of the new demands of our XXI st century school and students: be willing to try pedagogical approaches that integrate ICT into real teaching contexts is now a cause over which it is worth (re)thinking about.

## 5 FUTURE WORK

Because we performed our research with a convenience sample one cannot make generalizations. Thus, the results of our exploratory study reveal that it is worth to study more deeply our designed teaching strategy.

Future work should therefore focus on mechanisms that allow assessment both of the methodology used and the students-generated content in order to build a framework that helps to improve students' achievements with this SLE.